\documentclass[useAMS,usenatbib]{mn2e}
\usepackage{graphicx}

\title[Analytical studies on the SZ effect]{Analytical studies on the Sunyaev Zeldovich effect in the cluster of galaxies for three Lorentz frames}
\author[Satoshi Nozawa and Yasuharu Kohyama]{Satoshi Nozawa\thanks{E-mail: snozawa@josai.ac.jp} and Yasuharu Kohyama\thanks{c/o S. Nozawa, tentative address}\\
Josai Junior College, 1-1 Keyakidai, Sakado-shi, Saitama, 350-0295, Japan}

\begin{document}

\date{submitted: }

\pagerange{\pageref{firstpage}--\pageref{lastpage}} \pubyear{0000}

\maketitle

\label{firstpage}

\begin{abstract}
We study the Sunyaev-Zeldovich effect for clusters of galaxies.  The Boltzmann equations for the CMB photon distribution function are studied in three Lorentz frames.  We clarify the relations of the SZ effects among the different Lorentz frames.  We derive analytic expressions for the photon redistribution functions.  These formulas are applicable to the nonthermal electron distributions as well as the standard thermal distribution.  We show that the Fokker-Planck expansion of the Boltzmann equation can be expanded by the power series of the diffusion operator of the original Kompaneets equation.
\end{abstract}


\section{Introduction}

  The Sunyaev-Zeldovich (SZ) effect \citep{zeld69,suny80}, which arises from the Compton scattering of the cosmic microwave background (CMB) photons by hot electrons in clusters of galaxies (CG), provides a useful method for studies of cosmology.  For the reviews, for example, see \citet{birk99} and \citet{carl02}.  The original SZ formula has been derived from the Kompaneets equation \citep{komp56} in the non-relativistic approximation.  However, X-ray observations, for example, \citet{alle02} have revealed the existence of high-temperature CG such as $k_{B} T_{e} \simeq $20keV.  For such high-temperature CG, the relativistic corrections will become important.

  On the other hand, it has been known theoretically for some time that the relativistic corrections become significant at the short wave length region $\lambda < 1$mm.  In particular, the recent report on the first detection of the SZ effect at $\lambda < 650 \mu$m by the Herschel survey \citep{zemc10} seems to confirm the relativistic corrections.  Furthermore, new generation observations, for example, by \citet{plan11} are carrying out systematic studies of the precision measurements on the SZ effect.  Therefore, reliable theoretical studies on the relativistic SZ effect at short wave length will become extremely important for both existing and forthcoming observation projects.

  The theoretical studies on the relativistic corrections have been done by several groups.  \citet{wrig79} and \citet{reph95} have done the pioneering work to the SZ effect for the CG.  \citet{chal98} and \citet{itoh98} have adopted a relativistically covariant formalism to describe the Compton scattering process and have obtained higher-order relativistic corrections to the thermal SZ effect in the form of the Fokker-Planck expansion approximation.  \citet{noza98} have extended their method to the case where the CG is moving with a peculiar velocity $\vec{\beta}_{c}$ with respect to the CMB frame and have obtained the relativistic corrections to the kinematical SZ effect.  Their results were confirmed by \citet{chal99} and also by \citet{sazo98}.  \citet{itoh00} have also applied the covariant formalism to the polarization SZ effect \citep{suny81}.  The effect of the motion of the observer was also studied, for example, by \citet{chlu05} and \citet{noza05}.  The importance of the relativistic corrections is also exemplified through the possibility of directly measuring the cluster temperature using purely the SZ effect \citep{hans04}.  On the other hand, \citet{chlu12b} also studied the relativistic corrections on the SZ effect by calculating the Boltzmann equation in the CG frame and extending it to other frames.  They reported the importance of the separation of kinetic and scattering terms for the interpretation of future SZ data, where the retardation effect $(1 - \beta_{c} \mu_{c})$ gives a relevant correction.

  In the present paper, we give formal relations for the rate equations among different Lorentz frames, namely, for the CMB frame, CG frame and a general observer's (OBS) frame.  We clarify the relations of the SZ effects among the different Lorentz frames.  We derive analytic expressions for the photon redistribution functions.  These formulas are applicable to the nonthermal electron distributions as well as the standard thermal distribution.  We show that the Fokker-Planck expansion of the Boltzmann equation can be expanded by the power series of the diffusion operator of the original Kompaneets equation.

  The present paper is organized as follows.  In Section 2, we give formal relations for the rate equations among different Lorentz frames.  Starting from the Lorentz invariant Boltzmann equation for the photon-electron scattering in the CMB frame, we derive the relation formula between the CMB and CG frames.  Then, we extend it to the relation between the OBS and CG frames.  Integrating the rate equations along the line-of-sight, we formally show the Lorentz invariance of the SZ effect.  In Section 3, we derive the expression for the rate equation in the CG frame in terms of three redistribution functions $P_{\ell,c}(s_{c},\beta)$.  Then, we transform the rate equation in the CG frame to the CMB frame in Section 4.  We introduce new redistribution functions in the CMB frame with $P_{\ell,c}(s_{c},\beta)$.  In Section 5, we transform the rate equation in the CG frame to the OBS frame.  In Section 6, we derive analytic expressions for the redistribution functions.  With these formulas, 5-dimensional integrals for solving the rate equations are reduced to 2-dimensional integrals.  Concluding remarks will be given in Section 7.  We show the relations for the electron number densities among three Lorentz frames in Appendix A.  In Appendix B, we derive the rate equations in the CG and CMB frames with the Fokker-Planck expansion approximation for the thermal electron distribution.  We show that the Fokker-Planck expansion of the Boltzmann equation can be expanded by the power series of the diffusion operator of the original Kompaneets equation \citep{komp56}.  We also compare the present result with previous works.  Finally, we explore properties of the redistribution functions in Appendix C.

\section[]{General Formalism}

In the present section, we discuss the formal relations of the rate equations for the CMB photons among three different Lorentz frames, namely, the CMB frame, the CG frame and the OBS frame.  Recent work by \citet{chlu12b} also discussed the relations of the rate equations among different Lorentz frames by studying the Lorentz transformations of the optical depth.  And the results were calculated in the Fokker-Planck expansion approximation for the thermal electron distribution.  On the other hand, we show formal study based upon the general Lorentz invariant Boltzmann equation.  The results are presented without the Fokker-Planck expansion approximation.  The obtained expressions are applicable to arbitrary electron distributions.

  Let us start with the following Lorentz invariant expression of the Boltzmann equation for the electron-photon scattering in the CMB frame:
\begin{eqnarray}
\omega \frac{\partial n(\omega)}{\partial t} + \vec{k} \cdot \vec{\nabla} n(\omega) \nonumber  \\
&& \hspace{-36mm}
 = - \frac{2}{(2\pi)^{3}} \int \frac{d^{3}p}{E} \frac{d^{3}p^{\prime}}{E^{\prime}} \frac{d^{3}k^{\prime}}{2 \omega^{\prime}} \, \delta^{4}(p+k-p^{\prime}-k^{\prime}) \, \alpha^{2} \, \bar{X}  \nonumber  \\
&& \hspace{-33mm}
\times \left[ n(\omega) \left\{1 + n(\omega^{\prime}) \right\} f(E) - n(\omega^{\prime}) \left\{1 + n(\omega) \right\} f(E^{\prime}) \right] \, ,
\label{eq2-1}
\end{eqnarray}
where $\alpha$ is the fine structure constant, $k=(\omega,\vec{k})$ and $k^{\prime}=(\omega^{\prime},\vec{k}^{\prime})$ are the initial and final CMB photon momenta, respectively, and $p=(E,\vec{p})$ and $p^{\prime}=(E^{\prime},\vec{p}^{\, \prime})$ are the momenta for electrons.  In Eq.~(\ref{eq2-1}), $n(\omega)$ and $f(E)$ denote the distribution functions for the CMB photons and electrons in the CG, respectively, and $\bar{X}$ is the invariant transition probability of the Compton scattering, which will be discussed in Section 3.  In the present paper, we use the natural unit $\hbar=c=1$, unless otherwise stated explicitly.

  Assuming that the CMB photon distribution is spatially homogeneous, the space derivative term in the left hand side of Eq.~(\ref{eq2-1}) can be dropped.  In \citet{itoh98} and \citet{noza98}, the Boltzmann equation of Eq.~(\ref{eq2-1}) was calculated in the CMB frame under this assumption.

  On the other hand, the left hand side of Eq.~(\ref{eq2-1}) can be rewritten with variables in the CG frame as follows:
\begin{equation}
\omega \frac{\partial n(\omega)}{\partial t} + \vec{k} \cdot \vec{\nabla} n(\omega) = \omega_{c} \frac{\partial n_{c}(\omega_{c})}{\partial t_{c}} + \vec{k}_{c} \cdot \vec{\nabla}_{c} n_{c}(\omega_{c})  \, ,
\label{eq2-2}
\end{equation}
where $k_{c}=(\omega_{c}, \vec{k}_{c})$ denotes the momenta of the initial CMB photons in the CG frame, and $n_{c}(\omega_{c})$ is the CMB photon distribution function in the CG frame.  Let us consider that the CG is moving with a velocity $\vec{\beta}_{c}$ with respect to the CMB frame.  Then, the photon energies are related by the Lorentz transformation, $\omega_{c} = \gamma_{c} \left(1 - \beta_{c} \mu_{c} \right) \omega$, where $\gamma_{c} = 1/\sqrt{1 - \beta_{c}^{2}}$, and $\mu_{c} = \hat{\beta}_{c} \cdot \hat{k}$.  Assuming that the CMB photon distributions are spatially homogeneous in both frames, Eq.~(\ref{eq2-2}) becomes
\begin{equation}
\frac{d n(x)}{d t} = \gamma_{c} \left(1 - \beta_{c} \mu_{c} \right) \, \frac{d n_{c}(x_{c})}{d t_{c}}  \, ,
\label{eq2-3}
\end{equation}
where we introduced new variables $x_{c}\equiv\omega_{c}/k_{B}T_{\rm CMB}$ and $x\equiv\omega/k_{B}T_{\rm CMB}$.  Thus, Eq.~(\ref{eq2-3}) gives the relation for the rate equations between the CMB and CG frames.  Note that \citet{chlu12b} also derived Eq.~(\ref{eq2-3}) with different approach.  On the other hand, one can reverse the relation of Eq.~(\ref{eq2-3}) as
\begin{equation}
\frac{d n_{c}(x_{c})}{d t_{c}} = \gamma_{c} \left(1 + \beta_{c} \mu_{c}^{c} \right) \, \frac{d n(x)}{d t}  \, ,
\label{eq2-4}
\end{equation}
where $\mu_{c}^{c} = \hat{\beta}_{c} \cdot \hat{k}_{c}$.  As far as the CMB photon distributions are spatially homogenous in both frames, Eq.~(\ref{eq2-3}) and Eq.~(\ref{eq2-4}) are equivalent.  Finally, we rewrite Eq.~(\ref{eq2-3}) with new variables $\tau$ and $\tau_{c}$ as follows:
\begin{equation}
\frac{d n(x)}{d \tau} = \left(1 - \beta_{c} \mu_{c} \right) \frac{d n_{c}(x_{c})}{d \tau_{c}}  \, ,
\label{eq2-5}
\end{equation}
\begin{equation}
d\tau = N_{e} \sigma_{T} \, d t \, ,
\label{eq2-6}
\end{equation}
\begin{equation}
d\tau_{c} = N_{e}^{c} \sigma_{T} \, d t_{c} \, ,
\label{eq2-7}
\end{equation}
where $\sigma_{T}$ is the Thomson scattering cross section, and $N_{e}$ and $N_{e}^{c}$ are the electron number densities in the CMB and CG frames, respectively, which are related by $N_{e} = \gamma_{c} N_{e}^{c}$.

  Next, we extend Eq.~(\ref{eq2-2}) to the OBS frame moving with a velocity $\vec{\beta}_{o}$ with respect to the CMB frame.  Under the assumption of the spatial homogeneity for the CMB photon distributions, one has
\begin{equation}
\frac{d n_{o}(x_{o})}{d t_{o}} = \gamma_{o} \left(1 + \beta_{o} \mu_{o}^{o} \right) \, \frac{d n(x)}{d t}  \, ,
\label{eq2-8}
\end{equation}
where $\gamma_{o}=1/\sqrt{1-\beta_{o}^{2}}$, $\mu_{o}^{o} = \hat{\beta}_{o} \cdot \hat{k}_{o}$, $x_{o}=\omega_{o}/k_{B}T_{\rm CMB}$, and $(\omega_{o}, \vec{k}_{o})$ denotes the four momenta of the initial CMB photons in the OBS frame.  Inserting Eq.~(\ref{eq2-3}) into Eq.~(\ref{eq2-8}), one has
\begin{equation}
\frac{d n_{o}(x_{o})}{d t_{o}} = \gamma_{o} \gamma_{c} \left(1 + \beta_{o} \mu_{o}^{o} \right) \left(1 - \beta_{c} \mu_{c} \right) \, \frac{d n_{c}(x_{c})}{d t_{c}}  \, .
\label{eq2-9}
\end{equation}
\citet{chal02} discussed the transformation law for the photon propagation directions between two different Lorentz frames, which implies
\begin{equation}
\hat{k} = \left(\frac{ \hat{\beta}_{o} \cdot \hat{k}_{o} + \beta_{o} }{ 1 + \vec{\beta}_{o} \cdot \hat{k}_{o} } \right) \hat{\beta}_{o} + \, \frac{ \hat{k}_{o} - \hat{\beta}_{o} \cdot \hat{k}_{o} \, \hat{\beta}_{o} }{\gamma_{o} \left(1 + \vec{\beta}_{o} \cdot \hat{k}_{o} \right)} .
\label{eq2-10}
\end{equation}
Inserting Eq.~(\ref{eq2-10}) into Eq.~(\ref{eq2-9}), one obtains
\begin{equation}
\frac{d n_{o}(x_{o})}{d t_{o}} = \gamma_{o} \gamma_{c} \left(1 - \vec{\beta}_{o} \cdot \vec{\beta}_{c} \right) \left( 1 - \beta_{c} \mu_{c}^{o} + \beta_{o} \mu_{o}^{o} \right) \, \frac{d n_{c}(x_{c})}{d t_{c}}  \, ,
\label{eq2-11}
\end{equation}
where $\mu_{c}^{o} = \hat{\beta}_{c} \cdot \hat{k}_{o}$, and $\mathcal{O}(\beta_{c} \beta_{o}^{2}, \beta_{o} \beta_{c}^{2})$ terms were neglected.  Finally, we rewrite Eq.~(\ref{eq2-11}) with a new variable $\tau_{o}$ as
\begin{equation}
\frac{d n_{o}(x_{o})}{d \tau_{o}} = \left( 1 - \beta_{c} \mu_{c}^{o} + \beta_{o} \mu_{o}^{o} \right) \, \frac{d n_{c}(x_{c})}{d \tau_{c}}  \, ,
\label{eq2-12}
\end{equation}
\begin{equation}
d\tau_{o} = N_{e}^{o} \sigma_{T} \, dt_{o}  \, ,
\label{eq2-13}
\end{equation}
where $N_{e}^{o}$ is the electron number density in the OBS frame, which is related to $N_{e}^{c}$ by
\begin{equation}
N_{e}^{o} = \gamma_{o} \gamma_{c} \left(1 - \vec{\beta}_{o} \cdot \vec{\beta}_{c} \right) N_{e}^{c}  \, .
\label{eq2-14}
\end{equation}
In Appendix A, we give the explicit derivation of Eq.~(\ref{eq2-14}).  Thus, Eqs.~(\ref{eq2-5}) and (\ref{eq2-12}) summarize the formal relations for the rate equations among the CMB, CG and OBS frames.

  Finally, let us consider the integral of the rate equation in the CG frame along the line-of-sight $d\ell_{c}$.  One has
\begin{eqnarray}
\int_{L_{c}} d \ell_{c} \, \frac{d n_{c}(x_{c})}{cd t_{c}} = \int_{L} d \ell \, \gamma_{c} (1 - \beta_{c} \mu_{c}) \, \frac{d n_{c}(x_{c})}{cd t_{c}}
\nonumber  \\
&& \hspace{-48mm}
 = \int_{L} d \ell \, \frac{d n(x)}{cd t}   \, ,
\label{eq2-15}
\end{eqnarray}
where $d\ell_{c}=cdt_{c}$ and $d\ell=cdt$.  In deriving the first and last equalities in Eq.~(\ref{eq2-15}), we used the Lorentz transformations $dt_{c}=\gamma_{c}(1-\beta_{c} \mu_{c}) \, dt$ and Eq.~(\ref{eq2-3}), respectively.  Furthermore, Eq.~(\ref{eq2-15}) can be rewritten as
\begin{eqnarray}
\int_{L} d \ell \, \frac{d n(x)}{cd t} = \int_{L_{o}} d \ell_{o} \, \gamma_{o} (1 + \beta_{o} \mu_{o}^{o}) \, \frac{d n(x)}{cd t} \nonumber  \\
&& \hspace{-49mm}
= \int_{L_{o}} d \ell_{o} \, \frac{d n_{o}(x_{o})}{cd t_{o}}  \, ,
\label{eq2-16}
\end{eqnarray}
where $d\ell_{o}=cdt_{o}$, and we used the Lorentz transformations $dt = \gamma_{o} (1 + \beta_{o} \mu_{o}^{o}) \, dt_{o}$ and Eq.~(\ref{eq2-8}) in deriving the first and last equalities, respectively.  Thus, the integral of the rate equation along the line-of-sight (, which is the SZ effect) is Lorentz invariant.

\section{Calculations in the CG frame}

In the present section, we derive the rate equation in the CG frame.  Let us first define the kinematics in the CG frame.  The four momenta of the initial and final electrons are $p_{c}=(E_{c}, \vec{p}_{c})$ and $p_{c}^{\prime}=(E_{c}^{\prime},\vec{p}_{c}^{\, \prime})$, respectively, and $\vec{\beta}=\vec{p}_{c}/E_{c}$ is the velocity of the initial electron.  The four momenta of the initial and final photons are $k_{c}=(\omega_{c}, \vec{k}_{c})$ and $k_{c}^{\prime}=(\omega_{c}^{\prime},\vec{k}_{c}^{\prime})$, respectively.  According to \citet{noza09a}, the rate equation in the CG frame can be written in the Thomson approximation as follows:
\begin{eqnarray}
\frac{d n_{c}(x_{c})}{d \tau_{c}} = \frac{3}{64\pi^{2}} \int dp_{c} \, p_{c}^{2} \, p_{e,c}(E_{c})  \int d \Omega_{p_{c}} \int d \Omega_{k_{c}^{\prime}} \frac{1}{\gamma^{2}}
\nonumber \\
\times \frac{1}{1-\beta \mu} \left( \frac{\omega_{c}^{\prime}}{\omega_{c}} \right)^{2} \, \bar{X}_{A} \left[ n_{c}(x_{c}^{\prime}) - n_{c}(x_{c}) \right]  \, ,
\label{eq3-1-1}
\end{eqnarray}
\begin{equation}
\frac{\omega_{c}^{\prime}}{\omega_{c}} = \frac{1 - \beta \mu}{1 - \beta \mu^{\prime}}  \, ,
\label{eq3-1-2}
\end{equation}
where $\mu = \hat{\beta} \cdot \hat{k}_{c}$, $\mu^{\prime} = \hat{\beta} \cdot \hat{k}_{c}^{\prime}$, $\gamma = 1/\sqrt{1 - \beta^{2}}$, and the electron distribution function is normalized by $\int_{0}^{\infty} dp_{c} \, p_{c}^{2} \, p_{e,c}(E_{c})=1$.  In  Eq.~(\ref{eq3-1-1}), $\bar{X}_{A}$ is given by
\begin{equation}
\bar{X}_{A}  =  2 - \frac{2(1-{\rm cos}\Theta_{c})}
{\gamma^2(1-\beta\mu)(1-\beta\mu^{\prime})}
+ \frac{(1-{\rm cos} \Theta_{c})^2}{\gamma^4(1-\beta\mu)^2(1-\beta\mu^{\prime})^2} 
 \, ,
\label{eq3-1-3}
\end{equation}
\begin{equation}
{\rm cos}\Theta_{c} = \mu\mu^{\prime}+\sqrt{1-\mu^2} \sqrt{1-\mu^{\prime 2}}\cos(\phi_{k_{c}}-\phi_{k_{c}^{\prime}})  \, .
\label{eq3-1-4}
\end{equation}
It should be noted that the Thomson approximation used in deriving Eqs.~(\ref{eq3-1-1}) and (\ref{eq3-1-2}) is extremely good approximation for the CMB photon energies.

  Now, we expand the photon distribution functions in power series of $\beta_{c}$ up to $\mathcal{O}(\beta_{c}^{2})$ under the assumption $\beta_{c} \ll 1$.  One obtains the standard expressions \citep{chlu12a}:
\begin{eqnarray}
n_{c}(x_{c}) = n(x_{c}) + \frac{1}{6} \beta_{c}^{2} D_{x_{c}} ( D_{x_{c}} + 2 ) \, n(x_{c})  \nonumber  \\
&& \hspace{-43mm}
+ \beta_{c} P_{1}(\mu_{c}^{c}) \, D_{x_{c}} n(x_{c})  \nonumber  \\
&& \hspace{-43mm}
 + \frac{1}{3} \beta_{c}^{2} P_{2}(\mu_{c}^{c}) \, D_{x_{c}} (D_{x_{c}} - 1) \, n(x_{c})  \, ,
\label{eq3-1-5}
\end{eqnarray}
\begin{eqnarray}
n_{c}(x_{c}^{\prime}) = n(x_{c}^{\prime}) + \frac{1}{6} \beta_{c}^{2} D_{x_{c}^{\prime}} ( D_{x_{c}^{\prime}} + 2 ) \, n(x_{c}^{\prime})  \nonumber  \\
&& \hspace{-43mm}
+ \beta_{c} P_{1}(\mu_{c}^{c \prime}) \, D_{x_{c}^{\prime}} n(x_{c}^{\prime})    \nonumber  \\
&& \hspace{-43mm}
 + \frac{1}{3} \beta_{c}^{2} P_{2}(\mu_{c}^{c \prime}) \, D_{x_{c}^{\prime}} (D_{x_{c}^{\prime}} - 1) \, n(x_{c}^{\prime}) \, ,
\label{eq3-1-6}
\end{eqnarray}
where $\mu_{c}^{c} = \hat{\beta}_{c} \cdot \hat{k}_{c}$, $\mu_{c}^{c\prime} = \hat{\beta}_{c} \cdot \hat{k}_{c}^{\prime}$, $P_{\ell}(\mu_{c}^{c})$ and $P_{\ell}(\mu_{c}^{c \prime})$ are the Legendre polynomials of the $\ell$-th order, and
\begin{equation}
D_{z} \, \equiv \, z \, \frac{\partial}{\partial z}
\label{eq3-1-7}
\end{equation}
is the Lorentz invariant operator.  Choosing the direction of $\hat{\beta}_{c}$ along $z$-axis, one can reexpress $P_{\ell}(\mu_{c}^{c\prime})$ as
\begin{eqnarray}
P_{\ell}(\mu_{c}^{c\prime}) = \frac{4 \pi}{2 \ell + 1} \sum_{m} Y_{\ell,m}^{*}(\Theta_{c}, \phi_{k_{c}^{\prime}}) \, Y_{\ell,m}(\theta_{c}^{c}, \phi_{k_{c}})
\nonumber \\
&& \hspace{-63mm}
 = P_{\ell}({\rm cos} \Theta_{c}) \, P_{\ell}(\mu_{c}^{c})  \, ,
\label{eq3-1-8}
\end{eqnarray}
where $\phi_{k_{c}^{\prime}}$ and $\phi_{k_{c}}$ are the azimuthal angles of $\hat{k}_{c}^{\prime}$ and $\hat{k}_{c}$, respectively.  In deriving Eq.(\ref{eq3-1-8}), only $m=0$ term survived, because the photon distribution of $\hat{k}_{c}$ is axially symmetric for the z-axis.

  Then, we introduce a new variable $s_{c}$ by
\begin{equation}
x_{c}^{\prime} = e^{s_{c}} x_{c}  \, .
\label{eq3-1-9}
\end{equation}
Inserting Eqs.~(\ref{eq3-1-5}), (\ref{eq3-1-6}) and (\ref{eq3-1-8}) into Eq.~(\ref{eq3-1-1}), one finally obtains the rate equation in the CG frame as follows:
\begin{eqnarray}
\frac{d n_{c}(x_{c})}{d \tau_{c}} = \int_{-\infty}^{+\infty} ds_{c} \, P_{0,c}(s_{c}) \, \left[ n(e^{s_{c}}x_{c}) - n(x_{c}) \right] \nonumber  \\
&& \hspace{-75mm}
+ \frac{1}{6} \beta_{c}^{2} \int_{-\infty}^{+\infty} ds_{c} \, P_{0,c}(s_{c}) \, D_{x_{c}} (D_{x_{c}} + 2) \left[ n(e^{s_{c}}x_{c}) - n(x_{c}) \right]  \nonumber  \\
&& \hspace{-75mm}
+ \, \beta_{c} P_{1}(\mu_{c}^{c}) \left[ \int_{-\infty}^{+\infty} ds_{c} \, P_{1,c}(s_{c}) \, D_{x_{c}} n(e^{s_{c}}x_{c}) - D_{x_{c}} n(x_{c}) \right]
\nonumber  \\
&& \hspace{-75mm}
+ \, \frac{1}{3} \beta_{c}^{2} P_{2}(\mu_{c}^{c}) \left[ \int_{-\infty}^{+\infty} ds_{c} \, P_{2,c}(s_{c}) \, D_{x_{c}} (D_{x_{c}} - 1) \, n(e^{s_{c}}x_{c}) \right.  \nonumber  \\
&& \hspace{-33mm}
\left. - D_{x_{c}} (D_{x_{c}} - 1) \, n(x_{c}) \bigg] \right. \, ,
\label{eq3-1-10}
\end{eqnarray}
where
\begin{eqnarray}
\int_{-\infty}^{+\infty} ds_{c} \, P_{\ell,c}(s_{c}) \, n(e^{s_{c}}x_{c}) \nonumber  \\
&& \hspace{-45mm}
 = \frac{3}{64\pi^{2}} \int_{0}^{\infty} dp_{c} \, p_{c}^{2} \, p_{e,c}(E_{c}) \int d \Omega_{p_{c}} \int d \Omega_{k_{c}^{\prime}} \frac{1}{\gamma^{2}}  \nonumber \\
&& \hspace{-30mm}
\times  \frac{1-\beta \mu}{(1-\beta \mu^{\prime})^{2}} \, \bar{X}_{A} \, P_{\ell}({\rm cos} \Theta_{c}) \, n(x_{c}^{\prime})  \, .
\label{eq3-1-11}
\end{eqnarray}
In deriving Eq.~(\ref{eq3-1-10}), we used a familiar property of $P_{0,c}(s_{c})$,
\begin{equation}
\int_{-\infty}^{+\infty} ds_{c} \, P_{0,c}(s_{c}) \, = \, 1 \, .
\label{eq3-1-12}
\end{equation}

  In Appendix B1, we derive the expression for Eq.~(\ref{eq3-1-10}) in the Fokker-Planck expansion approximation for the thermal electron distribution.  In Appendix C, we show that the redistribution function $P_{\ell,c}(s_{c})$ satisfies the following useful relation:
\begin{equation}
P_{\ell,c}(s_{c}) = e^{3 s_{c}} \, P_{\ell,c}(-s_{c})
\label{eq3-1-13}
\end{equation}
for $\ell$=0, 1 and 2.  Because of the relation, one can use the same rate equation of Eq.~(\ref{eq3-1-10}) for the spectral intensity function $I(x_{c}) \equiv x_{c}^{3} n_{c}(x_{c})/2\pi^{2}$ except for replacing $e^{s_{c}}$ by $e^{-s_{c}}$ and $D_{x_{c}}$ by $D_{x_{c}}-3$.  Namely, one has
\begin{eqnarray}
\frac{d I_{c}(x_{c})}{d \tau_{c}} = \int_{-\infty}^{+\infty} ds_{c} \, P_{0,c}(s_{c}) \left[ I(e^{-s_{c}}x_{c}) - I(x_{c}) \right]
\nonumber  \\
&& \hspace{-75mm}
+ \frac{1}{6} \beta_{c}^{2} \int_{-\infty}^{+\infty} ds_{c} \, P_{0,c}(s_{c}) \, (D_{x_{c}} -3) (D_{x_{c}} -1) \nonumber  \\
&& \hspace{-40mm}
\times \left[ I(e^{-s_{c}}x_{c}) - I(x_{c}) \right]  \nonumber  \\
&& \hspace{-75mm}
+ \, \beta_{c} P_{1}(\mu_{c}^{c}) \left[ \int_{-\infty}^{+\infty} ds_{c} \, P_{1,c}(s_{c}) \, (D_{x_{c}} - 3) \, I(e^{-s_{c}}x_{c}) \right. \nonumber  \\
&& \hspace{-34mm}
- (D_{x_{c}} - 3) \, I(x_{c}) \bigg]
\nonumber  \\
&& \hspace{-75mm}
+ \, \frac{1}{3} \beta_{c}^{2} P_{2}(\mu_{c}^{c}) \left[ \int_{-\infty}^{+\infty} ds_{c} \, P_{2,c}(s_{c}) \, (D_{x_{c}} - 3) (D_{x_{c}} - 4) \right.  \nonumber  \\
&& \hspace{-58mm}
\times I(e^{-s_{c}}x_{c}) - (D_{x_{c}} - 3) (D_{x_{c}} - 4) \, I(x_{c}) \bigg]  \, .
\label{eq3-1-14}
\end{eqnarray}
It should be emphasized that the property of Eq.~(\ref{eq3-1-13}) for $\ell$=0, 1 and 2 are very remarkable features of the CG frame.

  Finally, integrating Eq.~(\ref{eq3-1-10}) over the phase space volume element $d^{3}x_{c}$, it is straightforward to verify the photon number conservation with Eqs.~(\ref{eq3-1-12}) and (\ref{eq3-1-13}).

\section[]{Transformation to the CMB frame}

In the present section, we rewrite Eq.~(\ref{eq3-1-10}) in terms of variables in the CMB frame.  Inserting the result into Eq.~(\ref{eq2-5}), one obtains the rate equation in the CMB frame.  The photon distribution functions are expanded up to $\mathcal{O}(\beta_{c}^{2})$ as follows:
\begin{eqnarray}
n(x_{c}) = n(x) + \frac{1}{6} \beta_{c}^{2} D_{x} (D_{x} + 2) \, n(x)  \nonumber  \\
&& \hspace{-40mm}
+ \beta_{c} P_{1}(\mu_{c}) \, D_{x} n(x)  \nonumber  \\
&& \hspace{-40mm}
 + \frac{1}{3} \beta_{c}^{2} P_{2}(\mu_{c}) \, D_{x} (D_{x} - 1) \, n(x)  \, ,
\label{eq4-1-1}
\end{eqnarray}
\begin{eqnarray}
n(e^{s_{c}} x_{c}) = n(e^{s_{c}} x) + \frac{1}{6} \beta_{c}^{2} D_{x} (D_{x} + 2) \, n(e^{s_{c}} x)  \nonumber  \\
&& \hspace{-44mm}
+ \beta_{c} P_{1}(\mu_{c}) \, D_{x} n(e^{s_{c}} x)    \nonumber  \\
&& \hspace{-44mm}
 + \frac{1}{3} \beta_{c}^{2} P_{2}(\mu_{c}) \, D_{x} (D_{x} - 1) \, n(e^{s_{c}} x) \, .
\label{eq4-1-2}
\end{eqnarray}
Similarly, one can expand
\begin{eqnarray}
\mu_{c}^{c} = \frac{\mu_{c} - \beta_{c}}{1 - \beta_{c}\mu_{c}}  = \mu_{c} + \frac{2}{3} \left\{ P_{2}(\mu_{c})-1 \right\} \beta_{c} + \mathcal{O}(\beta_{c}^{2})  \, .
\label{eq4-1-3}
\end{eqnarray}
Inserting Eqs.~(\ref{eq4-1-1})--(\ref{eq4-1-3}) into Eq.~(\ref{eq3-1-10}), one has
\begin{eqnarray}
\frac{d n_{c}(x, \mu_{c})}{d \tau_{c}} = \int_{-\infty}^{+\infty} ds \, P_{m}(s) \, \left[ n(e^{s}x) - n(x) \right]  \nonumber  \\
&& \hspace{-60mm}
+ \, \frac{1}{3} \beta_{c}^{2} \int_{-\infty}^{+\infty} ds \, P_{d}(s) \, D_{x}  (D_{x} + 2) \, n(e^{s} x) \nonumber  \\
&& \hspace{-60mm}
- \, \beta_{c} \, P_{1}(\mu_{c}) \int_{-\infty}^{+\infty} ds \, P_{d}(s) \, D_{x} \, n(e^{s} x)  \nonumber  \\
&& \hspace{-60mm}
+ \, \frac{1}{3} \beta_{c}^{2} \, P_{2}(\mu_{c})  \int_{-\infty}^{+\infty} ds \, P_{q}(s) \, D_{x} (D_{x} - 1) \, n(e^{s} x)  \, ,
\label{eq4-1-4}
\end{eqnarray}
where we introduced the following redistribution functions:
\begin{equation}
P_{m}(s) = P_{0,c}(s)  \, ,
\label{eq4-1-5}
\end{equation}
\begin{equation}
P_{d}(s) = P_{0,c}(s) - P_{1,c}(s)  \, ,
\label{eq4-1-6}
\end{equation}
\begin{equation}
P_{q}(s) = P_{2,c}(s) - 2 \, P_{1,c}(s) + P_{0,c}(s)  \, .
\label{eq4-1-7}
\end{equation}

  Before proceed to the final step in deriving the rate equation in the CMB frame, it should be noted as follows.  As indicated in Section 3, Eq.~(\ref{eq3-1-10}) conserved the photon number.  However, Eq.~(\ref{eq4-1-4}) which corresponds to Eq.~(25) of \cite{chlu12b} violates the photon number in the $\beta_{c}^{2}$ term.  This can be checked with the explicit forms for the $\mathcal{O}_{1}(D_{z})$ operator of Eqs.~(\ref{eqB-1-10})--(\ref{eqB-1-16}).  This illustrates that the photon number conservation is representation dependent.

  Inserting Eq.~(\ref{eq4-1-4}) into Eq.~(\ref{eq2-5}), one finally obtains the rate equation in the CMB frame as follows:
\begin{eqnarray}
\frac{d n(x)}{d \tau} = \int_{-\infty}^{+\infty} ds \, P_{m}(s) \, \left[ n(e^{s}x) - n(x) \right]  \nonumber  \\
&& \hspace{-57mm}
+ \, \frac{1}{3} \beta_{c}^{2} \int_{-\infty}^{+\infty} ds \, P_{d}(s) \, D_{x}  (D_{x} + 3) \, n(e^{s} x) \nonumber  \\
&& \hspace{-57mm}
- \, \beta_{c} \, P_{1}(\mu_{c}) \left[ \int_{-\infty}^{+\infty} ds \, P_{d}(s) \, D_{x} \, n(e^{s} x)  \right. \nonumber  \\
&& \hspace{-41mm}
\left. + \int_{-\infty}^{+\infty} ds \, P_{m}(s) \, \left\{ n(e^{s} x) - n(x) \right\}  \right]
\nonumber  \\
&& \hspace{-57mm}
+ \, \frac{1}{3} \beta_{c}^{2} \, P_{2}(\mu_{c}) \left[ \int_{-\infty}^{+\infty} ds \, P_{q}(s) \, D_{x} (D_{x} - 1) \, n(e^{s} x)  \right.  \nonumber  \\
&& \hspace{-37mm}
\left. + \, 2 \, \int_{-\infty}^{+\infty} ds \, P_{d}(s) \, D_{x} \, n(e^{s} x) \right]  \, .
\label{eq4-1-8}
\end{eqnarray}
In Appendix B2, we derive the expression for Eq.~(\ref{eq4-1-8}) in the Fokker-Planck expansion approximation for the thermal electron distribution and compare the result with previous works.  Note that one can use Eq.~(\ref{eq4-1-8}) for the spectral intensity function $I(x) \equiv x^{3} n(x)/2\pi^{2}$ except for replacing $e^{s}$ by $e^{-s}$ and $D_{x}$ by $D_{x}-3$.  This can be verified with Eq.~(\ref{eq3-1-13}).

  Finally, it should be noted that Eq.~(\ref{eq4-1-8}) satisfies the photon number conservation.  This can be checked by integrating Eq.~(\ref{eq4-1-8}) over the phase space volume element $d^{3}x$.

\section{Transformation to the general observer's frame}

In the present section, we now rewrite Eq.~(\ref{eq4-1-4}) in terms of variables in the OBS frame.  Inserting the result into Eq.~(\ref{eq2-12}), one obtains the expression for the rate equation in the OBS frame.  The photon distribution functions are expanded up to $\mathcal{O}(\beta_{o}^{2})$ under the assumption $\beta_{o} \ll 1$:
\begin{eqnarray}
n(x) = n(x_{o}) + \frac{1}{6} \beta_{o}^{2} D_{x_{o}} ( D_{x_{o}} + 2 ) \, n(x_{o})  \nonumber  \\
&& \hspace{-44mm}
+ \beta_{o} P_{1}(\mu_{o}^{o}) \, D_{x_{o}} n(x_{o})  \nonumber  \\
&& \hspace{-44mm}
 + \frac{1}{3} \beta_{o}^{2} P_{2}(\mu_{o}^{o}) \, D_{x_{o}} (D_{x_{o}} - 1) \, n(x_{o})  \, ,
\label{eq5-1}
\end{eqnarray}
\begin{eqnarray}
n(e^{s}x) = n(e^{s}x_{o}) + \frac{1}{6} \beta_{o}^{2} D_{x_{o}} ( D_{x_{o}} + 2 ) \, n(e^{s}x_{o})  \nonumber  \\
&& \hspace{-47mm}
+ \beta_{o} P_{1}(\mu_{o}^{o}) \, D_{x_{o}} n(e^{s}x_{o})  \nonumber  \\
&& \hspace{-47mm}
 + \frac{1}{3} \beta_{o}^{2} P_{2}(\mu_{o}^{o}) \, D_{x_{o}} (D_{x_{o}} - 1) \, n(e^{s}x_{o})  \, .
\label{eq5-2}
\end{eqnarray}
Similarly, $\beta_{c} P_{1}(\mu_{c})$ and $\beta_{c}^{2} P_{2}(\mu_{c})$ are transformed,
\begin{equation}
\beta_{c} P_{1}(\mu_{c}) = \beta_{c} P_{1}(\mu_{c}^{o}) + \vec{\beta}_{o} \cdot \vec{\beta}_{c} - \vec{\beta}_{o} \cdot \hat{k}_{o} \vec{\beta}_{c} \cdot \hat{k}_{o}  \, ,
\label{eq5-3}
\end{equation}
\begin{equation}
\beta_{c}^{2} P_{2}(\mu_{c}) = \beta_{c}^{2} P_{2}(\mu_{c}^{o}) \, .
\label{eq5-4}
\end{equation}
Inserting Eqs.~(\ref{eq5-1})--(\ref{eq5-4}) into Eq.~(\ref{eq4-1-4}), one finally obtains Eq.~(\ref{eq2-12}) as follows:
\begin{eqnarray}
\frac{d n_{o}(x_{o})}{d \tau_{o}} = \left[\frac{d n_{o}(x_{o})}{d \tau_{o}}\right]_{\beta_{c}} + \left[\frac{d n_{o}(x_{o})}{d \tau_{o}}\right]_{\beta_{o}}
 + \left[\frac{d n_{o}(x_{o})}{d \tau_{o}}\right]_{\beta_{o}\beta_{c}}  \, ,
\label{eq5-5}
\end{eqnarray}
where
\begin{eqnarray}
\left[\frac{d n_{o}(x_{o})}{d \tau_{o}}\right]_{\beta_{c}} = \int_{-\infty}^{+\infty} ds \, P_{m}(s) \, \left[ n(e^{s}x_{o}) - n(x_{o}) \right]  \nonumber  \\
&& \hspace{-75mm}
+ \frac{1}{3} \beta_{c}^{2} \int_{-\infty}^{+\infty} ds \, P_{d}(s) \, D_{x_{o}}  (D_{x_{o}} + 3) \, n(e^{s} x_{o}) \nonumber  \\
&& \hspace{-75mm}
- \, \beta_{c} \, P_{1}(\mu_{c}^{o}) \left[ \int_{-\infty}^{+\infty} ds \, P_{d}(s) \, D_{x_{o}} \, n(e^{s} x_{o})  \right. \nonumber  \\
&& \hspace{-57mm}
\left. + \int_{-\infty}^{+\infty} ds \, P_{m}(s) \, \left\{ n(e^{s}x_{o}) - n(x_{o}) \right\}  \right] \nonumber  \\
&& \hspace{-75mm}
+ \, \frac{1}{3} \beta_{c}^{2} \, P_{2}(\mu_{c}^{o}) \left[ \int_{-\infty}^{+\infty} ds \, P_{q}(s) \, D_{x_{o}} (D_{x_{o}} - 1) \, n(e^{s} x_{o})  \right.  \nonumber  \\
&& \hspace{-56mm}
\left. + \, 2 \, \int_{-\infty}^{+\infty} ds \, P_{d}(s) \, D_{x_{o}} \, n(e^{s} x_{o}) \right]  \, ,
\label{eq5-6}
\end{eqnarray}
\begin{eqnarray}
\left[\frac{d n_{o}(x_{o})}{d \tau_{o}}\right]_{\beta_{o}}  \nonumber  \\
&& \hspace{-25mm}
= \frac{1}{6} \beta_{o}^{2} \int_{-\infty}^{+\infty} ds \, P_{m}(s) \, D_{x_{o}}  (D_{x_{o}} + 4) \, \left[ n(e^{s}x_{o}) - n(x_{o}) \right] \nonumber  \\
&& \hspace{-25mm}
+ \, \beta_{o} \, P_{1}(\mu_{o}^{o}) \int_{-\infty}^{+\infty} ds \, P_{m}(s) \, \left(D_{x_{o}} + 1 \right) \, \left[ n(e^{s}x_{o}) - n(x_{o}) \right] \nonumber  \\
&& \hspace{-25mm}
+ \, \frac{1}{3} \beta_{o}^{2} \, P_{2}(\mu_{o}^{o}) \int_{-\infty}^{+\infty} ds \, P_{m}(s) \, D_{x_{o}} (D_{x_{o}} + 1) \nonumber  \\
&& \hspace{+15mm}
 \times \, \left[ n(e^{s}x_{o}) - n(x_{o}) \right]   \, ,
\label{eq5-7}
\end{eqnarray}
\begin{eqnarray}
\left[\frac{d n_{o}(x_{o})}{d \tau_{o}}\right]_{\beta_{o}\beta_{c}}  \nonumber  \\
&& \hspace{-27mm}
= - \frac{1}{3} \vec{\beta}_{o} \cdot \vec{\beta}_{c} \, \left[ \int_{-\infty}^{+\infty} ds \, P_{d}(s) \, D_{x_{o}} (D_{x_{o}}+3) \, n(e^{s}x_{o}) \right. \nonumber  \\
&& \hspace{-10mm}
\left. + \, \int_{-\infty}^{+\infty} ds \, P_{m}(s) \, D_{x_{o}} \, \left\{ n(e^{s}x_{o}) - n(x_{o}) \right\} \right]  \nonumber  \\
&& \hspace{-27mm}
- \, \left( \vec{\beta}_{o} \cdot \hat{k}_{o} \vec{\beta}_{c} \cdot \hat{k}_{o} - \frac{1}{3} \vec{\beta}_{o} \cdot \vec{\beta}_{c} \right) \left[ \int_{-\infty}^{+\infty} ds \, P_{d}(s) \, D_{x_{o}}^{2} n(e^{s}x_{o})  \right.   \nonumber  \\
&& \hspace{-18mm}
\left. + \, \int_{-\infty}^{+\infty} ds \, P_{m}(s) \, D_{x_{o}} \, \left\{ n(e^{s}x_{o}) - n(x_{o}) \right\} \right]  \, .
\label{eq5-8}
\end{eqnarray}

  The photon number conservation is manifest for Eq.~(\ref{eq5-5}) by the following reason.  Equation (\ref{eq5-6}) which is identical to Eq.~(\ref{eq4-1-8}) conserves the photon number.  The photon number conservation of Eq.~(\ref{eq5-7}) is checked with Eqs.~(\ref{eq3-1-12}) and (\ref{eq3-1-13}).  As for Eq.~(\ref{eq5-8}), the first two terms in the $[ \, ]$ brackets are also zeros by the similar reasons.  The last two terms vanish by the integral over the solid angles.

  Let us now study the expression of Eq.~(\ref{eq5-5}) in the limit $\vec{\beta}_{o} = \vec{\beta}_{c}$.  It is nontrivial whether the obtained expression has the correct form in the limit.  In this limit, one can show that $\mu_{o}^{o} = \mu_{c}^{o} = \mu_{c}^{c}$ and $N_{e}^{o}=N_{e}^{c}$.  Then, it is straightforward to show that Eq.~(\ref{eq5-5}) is reduced to Eq.~(\ref{eq3-1-10}).

  Finally, it should be remarked as follows.  In \citet{chlu12b}, it was suggested to use $\mu_{c}$ instead of $\mu_{c}^{o}$ in Eq.~(\ref{eq5-5}), because they are interested in $\mu_{c}$ for the study of the large-scale velocity fields.  These angles are related with Eq.~(\ref{eq2-10}) by
\begin{equation}
\mu_{c}^{o} = \mu_{c} + \beta_{o} \, \mu_{c} \, \mu_{o}^{o} - \vec{\beta}_{o} \cdot \hat{\beta}_{c} + \mathcal{O}(\beta_{o}^{2})  \,  .
\label{eq5-11}
\end{equation}
Inserting Eq.~(\ref{eq5-11}) into Eq.~(\ref{eq5-5}), one finds the following modifications.  As for Eq.~(\ref{eq5-6}), $P_{\ell}(\mu_{c}^{o})$ should be simply replaced by $P_{\ell}(\mu_{c})$ for $\ell=0,1,2$.  There are no modifications to Eq.~(\ref{eq5-7}).  On the other hand, Eq.~(\ref{eq5-8}) is modified as
\begin{eqnarray}
\left[\frac{d n_{o}(x_{o})}{d \tau_{o}}\right]_{\beta_{o}\beta_{c}}  \nonumber  \\
&& \hspace{-27mm}
= - \frac{1}{3} \vec{\beta}_{o} \cdot \vec{\beta}_{c} \, \left[ \int_{-\infty}^{+\infty} ds \, P_{d}(s) \, D_{x_{o}} (D_{x_{o}}+1) \, n(e^{s}x_{o}) \right. \nonumber  \\
&& \hspace{-13mm}
\left. + \, \int_{-\infty}^{+\infty} ds \, P_{m}(s) \, (D_{x_{o}}-2) \, \left\{ n(e^{s}x_{o}) - n(x_{o}) \right\} \right]  \nonumber  \\
&& \hspace{-27mm}
- \, \left( \vec{\beta}_{o} \cdot \hat{k} \vec{\beta}_{c} \cdot \hat{k} - \frac{1}{3} \vec{\beta}_{o} \cdot \vec{\beta}_{c} \right)  \nonumber  \\
&& \hspace{-25mm}
\times \left[ \, \int_{-\infty}^{+\infty} ds \, P_{d}(s) \, D_{x_{o}}(D_{x_{o}}+1) n(e^{s}x_{o})  \right.   \nonumber  \\
&& \hspace{-23mm}
\left. + \, \int_{-\infty}^{+\infty} ds \, P_{m}(s) \, (D_{x_{o}}+1) \, \left\{ n(e^{s}x_{o}) - n(x_{o}) \right\} \right]  \, .
\label{eq5-13}
\end{eqnarray}
It is clear that the first term in the $\vec{\beta}_{o} \cdot \vec{\beta}_{c}$ terms does not conserve the photon number unless $\vec{\beta}_{o} \cdot \vec{\beta}_{c}=0$, which again illustrates that the photon number conservation is representation dependent.

\section{Analytic Expressions for the redistribution functions}

For the practical calculation of the rate equation, Eq.~(\ref{eq3-1-10}) can be solved by calculating 5-dimensional integrals of Eq.~(\ref{eq3-1-11}).  For the present computer technologies, the direct numerical integration can be done within a reasonable CPU time.  However, it is important to derive analytic expressions not only to reduce the CPU time of the computer, but also to investigate properties of functions in the equation.  The obtained analytic expression will be useful to analyze observational data.

  According to \citet{noza09a}, we introduce the following Lorentz transformations for the photon angles from the CG frame to the electron rest (ER) frame:
\begin{equation}
\mu = \frac{- \mu_{0} + \beta}{1 - \beta \mu_{0}} \, ,
\label{eq6-1-1}
\end{equation}
\begin{equation}
\mu^{\prime} = \frac{- \mu_{0}^{\prime} + \beta}{1 - \beta \mu_{0}^{\prime}}  \, ,
\label{eq6-1-2}
\end{equation}
where $\mu_{0} = \hat{\beta} \cdot \hat{k}_{0}$ and $\mu_{0}^{\prime} = \hat{\beta} \cdot \hat{k}_{0}^{\prime}$.  Then, Eqs.~(\ref{eq3-1-3}) and (\ref{eq3-1-4}) become
\begin{equation}
\bar{X}_{A,0} = \frac{4}{3} \, P_{0}({\rm cos} \Theta_{0}) + \frac{2}{3} \, P_{2}({\rm cos} \Theta_{0})  \, ,
\label{eq6-1-3}
\end{equation}
\begin{equation}
{\rm cos}\Theta_{0} = \mu_{0}\mu_{0}^{\prime}+\sqrt{1-\mu_{0}^2} \sqrt{1-\mu_{0}^{\prime 2}}\cos(\phi_{k_{c}}-\phi_{k_{c}^{\prime}})  \, .
\label{eq6-1-4}
\end{equation}
Similarly, $P_{\ell}({\rm cos} \Theta_{c})$ can be reexpressed by
\begin{equation}
P_{0}({\rm cos}\Theta_{c}) = P_{0}({\rm cos}\Theta_{0}) = 1 \, ,
\label{eq6-1-5}
\end{equation}
\begin{equation}
P_{1}({\rm cos}\Theta_{c}) = 1 - a + a \, P_{1}({\rm cos}\Theta_{0})  \, ,
\label{eq6-1-6}
\end{equation}
\begin{eqnarray}
P_{2}({\rm cos}\Theta_{c}) = 1 - 3a + 2a^{2} + 3a(1-a) \, P_{1}({\rm cos}\Theta_{0})
\, \nonumber \\
&& \hspace{-39mm}
 + a^{2} P_{2}({\rm cos}\Theta_{0})  \, ,
\label{eq6-1-7}
\end{eqnarray}
where
\begin{equation}
a = \frac{1}{\gamma^{2}(1-\beta \mu_{0})(1-\beta \mu_{0}^{\prime})} \, .
\label{eq6-1-8}
\end{equation}
Inserting Eqs.~(\ref{eq6-1-1}) and (\ref{eq6-1-2}) into Eq.~(\ref{eq3-1-2}), one has
\begin{equation}
e^{s_{c}} = \frac{\omega_{c}^{\prime}}{\omega_{c}} = \frac{1-\beta\mu_{0}^{\prime}}{1-\beta\mu_{0}}  \, .
\label{eq6-1-9}
\end{equation}

  One can then express $P_{\ell,c}(s_{c})$ in Eq.~(\ref{eq3-1-10}) by
\begin{equation}
P_{\ell,c}(s_{c}) = \int_{\beta_{\rm min}}^{1}  d \beta \, \beta^{2} \gamma^{5} \tilde{p}_{e,c}(\beta) \, P_{\ell,c}(s_{c},\beta)  \, ,
\label{eq6-1-10}
\end{equation}
where $\beta_{\rm min} =(1 - e^{-|s_{c}|})/(1 +e^{-|s_{c}|})$, $\tilde{p}_{e,c}(\beta) \equiv m^{3} p_{e,c}(E_{c})$, $m$ is the electron rest mass.  And
\begin{eqnarray}
P_{\ell,c}(s_{c},\beta) =  P_{\ell 0,c}(s_{c}, \beta) + P_{\ell 2,c}(s_{c},\beta)  \nonumber  \\
&& \hspace{-60mm}
= \frac{3 e^{s_{c}}}{64 \pi^{2} \beta \gamma^{4}} \int_{\mu_{1}(s_{c})}^{\mu_{2}(s_{c})} d\mu_{0} \frac{1}{(1-\beta \mu_{0})^{2}} \int_{0}^{2 \pi} d \phi_{k_{c}}  \int_{0}^{2 \pi} d \phi_{k_{c}^{\prime}}  \nonumber  \\
&& \hspace{-45mm}
\times   P_{\ell}({\rm cos} \Theta_{c}) \, \left\{ \frac{4}{3} \, P_{0}({\rm cos} \Theta_{0}) + \frac{2}{3} \, P_{2}({\rm cos} \Theta_{0})  \right\}  \, , 
\label{eq6-1-11}
\end{eqnarray}
where
\begin{eqnarray}
\mu_1(s_{c}) = \left\{
\begin{array}{ll}
-1 &\quad  {\rm for} \, \, \, s_{c} < 0 \\
{[1-e^{-s_{c}}(1+\beta)]/\beta} &\quad {\rm for} \, \, \, s_{c} > 0
\end{array}
\right.  \, ,
\label{eq6-1-12}
\end{eqnarray}
\begin{eqnarray}
\mu_2(s_{c}) = \left\{
\begin{array}{ll}
{[1-e^{-s_{c}}(1-\beta)]/\beta} &\quad {\rm for} \, \, \, s_{c} < 0 \\
1 &\quad  {\rm for} \, \, \, s_{c} > 0 
\end{array}
\right. \, .
\label{eq6-1-13}
\end{eqnarray}

  After lengthy but straightforward calculation, one finally obtains the following results:
\begin{equation}
P_{00,c}(s_{c},\beta) = \frac{e^{3s_{c}/2}}{2 \beta^{2} \gamma^{2}} \left[ \beta \left( {\rm cosh} \frac{s_{c}}{2} - \frac{1}{\beta} \, {\rm sinh} \frac{|s_{c}|}{2} \right) \right]  \, ,
\label{eq6-1-14}
\end{equation}
\begin{eqnarray}
P_{02,c}(s_{c},\beta) = \frac{1}{2} \, P_{00,c}(s_{c},\beta)  \nonumber  \\
&& \hspace{-45mm}
+ \frac{e^{3s_{c}/2}}{4 \beta^{2} \gamma^{2}} \left[ \frac{3}{\beta \gamma^{2}} \left( {\rm cosh}^{2} \frac{s_{c}}{2} - \frac{1}{\beta^{2}} \, {\rm sinh}^{2} \frac{ |s_{c}|}{2} \right) \right.  \nonumber  \\
&& \hspace{-27mm}
\times \left( 3 \, {\rm cosh} \frac{s_{c}}{2} - \frac{2}{\beta} \, {\rm sinh} \frac{|s_{c}|}{2} \right)  \nonumber  \\
&& \hspace{-33mm}
+ \frac{9}{\beta^{3} \gamma^{4}} \left({\rm cosh} \frac{s_{c}}{2} - \frac{1}{\beta} \, {\rm sinh} \frac{|s_{c}|}{2} \right) {\rm cosh}^{2} \frac{s_{c}}{2} \nonumber  \\
&& \hspace{-33mm}
\left. - \frac{3}{2\beta^{4} \gamma^{2}} \left(3 - \beta^{2} \right) \left(\lambda_{\beta} - |s_{c}| \right) \, {\rm cosh} \frac{s_{c}}{2} \right]   \, ,
\label{eq6-1-15}
\end{eqnarray}
where cosh$z=(e^{z}+e^{-z})/2$, sinh$z=(e^{z}-e^{-z})/2$ and $\lambda_{\beta}=\ln[(1+\beta)/(1-\beta)]$.  Note that Eq.~(\ref{eq6-1-14}) is the familiar expression in the isotropic scattering approximation.  On the other hand, Eq.~(\ref{eq6-1-15}) gives the anisotropic scattering contribution.  Similarly,
\begin{eqnarray}
P_{10,c}(s_{c},\beta)  \nonumber  \\
&& \hspace{-22mm}
= \frac{e^{3s_{c}/2}}{2 \beta^{2} \gamma^{2}} \left[ \frac{1}{3} \beta \left( {\rm cosh} \frac{s_{c}}{2} - \frac{1}{\beta} \, {\rm sinh} \frac{|s_{c}|}{2} \right)^{3} \right.  \nonumber  \\
&& \hspace{-11mm}
- \frac{1}{\beta \gamma^{2}} \left( {\rm cosh} \frac{s_{c}}{2} - \frac{1}{\beta} \, {\rm sinh} \frac{|s_{c}|}{2} \right) {\rm sinh}^{2} \frac{|s_{c}|}{2} \Bigg]  \, ,
\label{eq6-1-16}
\end{eqnarray}
\begin{eqnarray}
P_{12,c}(s_{c},\beta) = \frac{1}{2} \, P_{10,c}(s_{c},\beta)  \nonumber  \\
&& \hspace{-45mm}
+ \frac{e^{3s_{c}/2}}{4 \beta^{2} \gamma^{2}} \left[ \frac{3}{\beta \gamma^{2}} \left( {\rm cosh}^{2} \frac{s_{c}}{2} - \frac{1}{\beta^{2}} \, {\rm sinh}^{2} \frac{ |s_{c}|}{2} \right)  \right.  \nonumber  \\
&& \hspace{-27mm}
\times  \left( 5 \, {\rm cosh} \frac{s_{c}}{2} - \frac{2}{\beta} \, {\rm sinh} \frac{|s_{c}|}{2} \right)  \nonumber  \\
&& \hspace{-33mm}
+ \frac{1}{\beta^{3} \gamma^{4}} \left( {\rm cosh} \frac{s_{c}}{2} - \frac{1}{\beta} \, {\rm sinh} \frac{ |s_{c}|}{2} \right) \left( 82 \, {\rm cosh}^{2} \frac{s_{c}}{2}  \right.  \nonumber  \\
&& \hspace{-22mm}
\left. + \frac{25}{\beta} \, {\rm cosh} \frac{s_{c}}{2} \, {\rm sinh} \frac{ |s_{c}| }{2} - \frac{20}{\beta^{2}} \, {\rm sinh}^{2} \frac{ |s_{c}|}{2} \right)  \nonumber  \\
&& \hspace{-33mm}
+ \frac{75}{\beta^{5} \gamma^{6}} \left( {\rm cosh} \frac{s_{c}}{2} - \frac{1}{\beta} \, {\rm sinh} \frac{ |s_{c}|}{2} \right) {\rm cosh}^{2} \frac{s_{c}}{2}  \nonumber  \\
&& \hspace{-33mm}
- \frac{3}{2 \beta^{4} \gamma^{2}} \left(\lambda_{\beta} - |s_{c}| \right) \left(3 - \beta^{2} \right) {\rm cosh} \frac{s_{c}}{2}  \nonumber  \\
&& \hspace{-33mm}
- \frac{3}{2 \beta^{6} \gamma^{4}} \left(\lambda_{\beta} - |s_{c}| \right) \left\{ 3 \left(5 - \beta^{2} \right) {\rm cosh} \frac{s_{c}}{2} \right.  \nonumber  \\
&& \hspace{-6mm}
\left. + 2 \left(5 - 3 \beta^{2} \right) {\rm cosh}^{3} \frac{s_{c}}{2} \right\} \bigg]  \, ,
\label{eq6-1-17}
\end{eqnarray}
\begin{eqnarray}
P_{20,c}(s_{c},\beta)  \nonumber  \\
&& \hspace{-22mm}
= \frac{e^{3s_{c}/2}}{2 \beta^{2} \gamma^{2}} \left[ \frac{1}{5} \beta \left({\rm cosh} \frac{s_{c}}{2} - \frac{1}{\beta} \, {\rm sinh} \frac{|s_{c}|}{2} \right)^{5}  \right.  \nonumber  \\
&& \hspace{-11mm}
- \frac{2}{\beta \gamma^{2}} \left( {\rm cosh} \frac{s_{c}}{2} - \frac{1}{\beta} \, {\rm sinh} \frac{|s_{c}|}{2} \right)^{2} {\rm cosh} \, \frac{s_{c}}{2} \, {\rm sinh}^{2} \frac{|s_{c}|}{2}  \nonumber  \\
&& \hspace{-11mm}
+ \frac{1}{\beta^{3} \gamma^{4}} \left( {\rm cosh} \frac{s_{c}}{2} - \frac{1}{\beta} \, {\rm sinh} \frac{|s_{c}|}{2} \right) {\rm sinh}^{4} \frac{|s_{c}|}{2}   \Bigg]  \, ,
\label{eq6-1-18}
\end{eqnarray}
\begin{eqnarray}
P_{22,c}(s_{c},\beta) = \frac{1}{2} \, P_{20,c}(s_{c},\beta)  \nonumber  \\
&& \hspace{-45mm}
+ \frac{e^{3s_{c}/2}}{4 \beta^{2} \gamma^{2}} \left[ \frac{6}{\beta \gamma^{2}} \left( {\rm cosh}^{2} \frac{s_{c}}{2} - \frac{1}{\beta^{2}} \, {\rm sinh}^{2} \frac{ |s_{c}|}{2} \right)^{2}  \right.  \nonumber  \\
&& \hspace{-27mm}
\times  \left( 3 \, {\rm cosh} \frac{s_{c}}{2} - \frac{1}{\beta} \, {\rm sinh} \frac{|s_{c}|}{2} \right)  \nonumber  \\
&& \hspace{-33mm}
+ \frac{3}{\beta^{3} \gamma^{4}} \left( {\rm cosh}^{2} \frac{s_{c}}{2} - \frac{1}{\beta^{2}} \, {\rm sinh}^{2} \frac{ |s_{c}|}{2} \right)  \nonumber  \\
&& \hspace{-25mm}
\times \left( 94 \, {\rm cosh}^{3} \frac{s_{c}}{2} - \frac{51}{\beta} \, {\rm cosh}^{2} \frac{s_{c}}{2} \, {\rm sinh} \frac{ |s_{c}| }{2}  \right.   \nonumber  \\
&& \hspace{-23mm}
\left. - \frac{54}{\beta^{2}} \, {\rm cosh} \frac{s_{c}}{2} \, {\rm sinh}^{2} \frac{ |s_{c}|}{2} + \frac{18}{\beta^{3}} \, {\rm sinh}^{3} \frac{ |s_{c}|}{2} \right)  \nonumber  \\
&& \hspace{-33mm}
+ \frac{3}{4 \beta^{5} \gamma^{6}} \left( {\rm cosh} \frac{s_{c}}{2} - \frac{1}{\beta} \, {\rm sinh} \frac{ |s_{c}|}{2} \right)  \nonumber  \\
&& \hspace{-24mm}
\times \left( 32 \, {\rm cosh}^{2} \frac{s_{c}}{2} - \frac{59}{\beta} \, {\rm cosh} \frac{s_{c}}{2} \, {\rm sinh} \frac{ |s_{c}| }{2}  \right.   \nonumber  \\
&& \hspace{-22mm}
\left. + \frac{40}{\beta^{2}} \, {\rm sinh}^{2} \frac{ |s_{c}|}{2} \right) {\rm sinh}^{2} \frac{ |s_{c}|}{2}  \nonumber  \\
&& \hspace{-33mm}
+ \frac{15}{4 \beta^{5} \gamma^{6}} \left( {\rm cosh} \frac{s_{c}}{2} - \frac{1}{\beta} \, {\rm sinh} \frac{ |s_{c}|}{2} \right)  \nonumber  \\
&& \hspace{-24mm}
\times \left( 211 \, {\rm cosh}^{2} \frac{s_{c}}{2} + \frac{49}{\beta} \, {\rm cosh} \frac{s_{c}}{2} \, {\rm sinh} \frac{ |s_{c}| }{2}  \right.   \nonumber  \\
&& \hspace{-22mm}
\left. - \frac{161}{\beta^{2}} \, {\rm sinh}^{2} \frac{ |s_{c}|}{2} \right) {\rm cosh}^{2} \frac{ |s_{c}|}{2}  \nonumber  \\
&& \hspace{-33mm}
+ \frac{2205}{4 \beta^{7} \gamma^{8}} \left( {\rm cosh} \frac{s_{c}}{2} - \frac{1}{\beta} \, {\rm sinh} \frac{ |s_{c}|}{2} \right) {\rm cosh}^{4} \frac{s_{c}}{2}  \nonumber  \\
&& \hspace{-33mm}
- \frac{3}{2 \beta^{4} \gamma^{2}} \left(\lambda_{\beta} - |s_{c}| \right) \left(3 - \beta^{2} \right) {\rm cosh} \frac{s_{c}}{2}  \nonumber  \\
&& \hspace{-33mm}
- \frac{9}{2 \beta^{6} \gamma^{4}} \left(\lambda_{\beta} - |s_{c}| \right) \left\{ 3 \left(5 - \beta^{2} \right) {\rm cosh} \frac{s_{c}}{2} \right.  \nonumber  \\
&& \hspace{-6mm}
\left. + 2 \left(5 - 3 \beta^{2} \right) {\rm cosh}^{3} \frac{s_{c}}{2} \right\}  \nonumber  \\
&& \hspace{-33mm}
- \frac{45}{8 \beta^{8} \gamma^{6}} \left(\lambda_{\beta} - |s_{c}| \right) \left\{ 3 \left(7 - \beta^{2} \right) {\rm cosh} \frac{s_{c}}{2} \right.  \nonumber  \\
&& \hspace{-6mm}
\left.  + 4 \left(7 - 3 \beta^{2} \right) {\rm cosh}^{3} \frac{s_{c}}{2} \right\} \Bigg]  \, .
\label{eq6-1-19}
\end{eqnarray}
Thus, Eqs.(\ref{eq6-1-14})--(\ref{eq6-1-19}) summarize the analytic expressions for the redistribution functions in the CG frame.  Analytic expressions for the CMB and OBS frames are obtained with Eqs.~(\ref{eq4-1-5})--(\ref{eq4-1-7}).   With these formulas, 5-dimensional integrals for the rate equations were reduced to 2-dimensional integrals.

  Finally, it should be emphasized that the present analytic formulas have advantages compared with previous analytic formulas \citep{itoh98, chlu12b} and/or the analytic fitting formulas \citep{noza00} of the numerical calculation.  The present formulas were derived directly from the underling scattering physics without introducing any expansion approximations.  Therefore, they are free from the artificial oscillations due to the power series expansions.  Furthermore, the previous analytic formulas are restricted to the cases of the thermal electron distribution.  On the other hand, the present formulas are applicable to nonthermal electron distributions as well as the standard thermal distribution.  Thus, the present formulas provide us a wider applicability for the analysis of the observational data.

\section{Conclusions}

We studied the Sunyaev-Zeldovich effect for the clusters of galaxies in the Thomson approximation.  In Section 2, we started with the Lorentz invariant expression of the Boltzmann equation for the electron-photon scattering in the CMB frame.  Assuming that the CMB photon distributions are spatially homogeneous both in the CMB and CG frames, we derived the formal relation for the rate equations between the CMB and CG frames, which agreed with \citet{chlu12b}.  We extended the formula to the relation between the CG and OBS frames.  The formal relations for the rate equations among three Lorentz frames have been established.  By integrating the rate equations along the line-of-sight, we showed the Lorentz invariance of the SZ effect (the number of scatterings for a given photon frequency).  Thus, we clarified the relations of the SZ effects among the different Lorentz frames.

  In Section 3, we investigated the rate equation for the photon distribution function in the CG frame.  We introduced three redistribution functions $P_{\ell,c}(s_{c})$ ($\ell=0,1,2$) and expressed the rate equation in terms of these functions.    We also reexpressed the rate equation in the operator representation in Appendix B1.  We showed the rate equation in the Fokker-Planck expansion approximation for the thermal electron distribution in terms of the diffusion operator $\Delta_{z}$ of the original Kompaneets equation \citep{komp56}.

  In Section 4, we transformed the rate equation in the CG frame to the CMB frame with the formula established in Section 2.  We calculated the rate equation in the Fokker-Planck expansion approximation for the thermal electron distribution in Appendix B2.  We compared the present result with the previous works such as \citet{noza98} and \citet{chlu12b}.  We found that the present work agreed with these works completely when we multiplied the factor $1/\gamma_{c}$ to the result  of \citet{noza98}.

  In Section 5, we transformed the rate equation in the CG frame to the OBS frame.  The obtained expression had advantages compared with previous works.  First, the photon number conservation was manifestly realized in the present form.  Secondly, in the limit of $\vec{\beta}_{o} = \vec{\beta}_{c}$ the present expression was correctly reduced to the form in the CG frame.

  Finally, we derived analytic expressions for the redistribution functions $P_{\ell,c}(s_{c},\beta)$ for $\ell=0,1,2$ in Section 6.  The present formulas are applicable to nonthermal electron distributions as well as the standard thermal distribution.  With these formulas, 5-dimensional integrals for solving the rate equations were reduced to 2-dimensional integrals.  These analytic expressions are useful for the analysis of the observational data.

\section*{Acknowledgments}

We wish to acknowledge Naoki Itoh for enlightening us on this subject and also for useful suggestions.  We also thank Jens Chluba and our referee for valuable suggestions.

\appendix

\section[]{Electron number densities}

We study relations of the electron number densities ($N_{e}^{c}$, $N_{e}$, $N_{e}^{o}$) among three Lorentz frames.
We define the isotropic electron distribution function $f_{c}(E_{c})$ in the CG frame as
\begin{equation}
2 \int \frac{d^{3}p_{c}}{(2\pi)^{3}} \, f_{c}(E_{c}) = N_{e}^{c}  \, .
\label{eqA-1}
\end{equation}
Similarly, $f(E)$ in the CMB frame and $f_{o}(E_{o})$ in the OBS frame are defined by
\begin{equation}
2 \int \frac{d^{3}p}{(2\pi)^{3}} \, f(E) = N_{e}  \, ,
\label{eqA-2}
\end{equation}
\begin{equation}
2 \int \frac{d^{3}p_{o}}{(2\pi)^{3}} \, f_{o}(E_{o}) = N_{e}^{o}  \, .
\label{eqA-3}
\end{equation}
The CG and OBS are moving with velocities $\vec{\beta}_{c}$ and $\vec{\beta}_{o}$, respectively, with respect to the CMB frame.

  We study the relation between $N_{e}^{o}$ and $N_{e}^{c}$.  The energies and momenta are related by the Lorentz transformations,
\begin{equation}
E_{o} = \gamma_{o} \left(1 - \frac{\vec{\beta}_{o} \cdot \vec{p}}{E} \right) E \, ,
\label{eqA-4}
\end{equation}
\begin{equation}
d^{3}p_{o} = \gamma_{o} \left(1 - \frac{\vec{\beta}_{o} \cdot \vec{p}}{E} \right) d^{3} p \, .
\label{eqA-5}
\end{equation}
In order to Lorentz boost the right hand side of Eq.~(\ref{eqA-5}) to the CG frame, we decompose $\vec{p}$ as
\begin{equation}
\vec{p} = p_{\parallel} \hat{\beta}_{c} + p_{\perp} \hat{\beta}_{\perp}  \, ,
\label{eqA-6}
\end{equation}
where $p_{\parallel}=\hat{\beta}_{c} \cdot \vec{p}$ and $p_{\perp} \hat{\beta}_{\perp} = \vec{p} - (\hat{\beta}_{c} \cdot \vec{p}) \hat{\beta}_{c}$.  The vectors $\hat{\beta}_{c}$ and $\hat{\beta}_{\perp}$ satisfy the conditions: $\hat{\beta}_{c} \cdot \hat{\beta}_{\perp} = 0$ and $\hat{\beta}_{c} \cdot \hat{\beta}_{c} = \hat{\beta}_{\perp} \cdot \hat{\beta}_{\perp} = 1$.  Then, one obtains
\begin{equation}
\frac{\vec{p}}{E}  = \frac{ \left(\displaystyle{ \frac{\hat{\beta}_{c} \cdot \vec{p}_{c}}{E_{c}}} + \beta_{c} \right) \hat{\beta}_{c} }{\displaystyle{1 + \frac{\vec{\beta}_{c} \cdot \vec{p}_{c}}{E_{c}}}} + \frac{ \displaystyle{ \frac{\vec{p}_{c}}{E_{c}} - \frac{\hat{\beta}_{c} \cdot \vec{p}_{c} \hat{\beta}_{c}}{E_{c}} } }{\gamma_{c} \left(\displaystyle{1 + \frac{\vec{\beta}_{c} \cdot \vec{p}_{c}}{E_{c}}}\right) }  \, .
\label{eqA-7}
\end{equation}
Note that Eq.~(\ref{eqA-7}) is reduced to Eq.~(\ref{eq2-10}) in the massless limit.  Inserting Eq.~(\ref{eqA-7}) and
\begin{equation}
d^{3}p =  \gamma_{c} \left(1 + \frac{\vec{\beta}_{c} \cdot \vec{p}_{c}}{E_{c}} \right)  d^{3}p_{c}
\label{eqA-8}
\end{equation}
into Eq.~(\ref{eqA-5}), one has
\begin{eqnarray}
d^{3}p_{o} = \gamma_{o} \gamma_{c} \bigg[1 - \vec{\beta}_{o} \cdot \vec{\beta}_{c} + \bigg\{ \vec{\beta}_{c} - \vec{\beta}_{o} \cdot \hat{\beta}_{c} \hat{\beta}_{c}\nonumber  \\
&& \hspace{-40mm}
 - \frac{1}{\gamma_{c}} \left(\vec{\beta}_{o} - \vec{\beta}_{o} \cdot \hat{\beta}_{c} \hat{\beta}_{c} \right) \bigg\} \cdot \frac{\vec{p}_{c}}{E_{c}} \bigg] d^{3} p_{c} \, .
\label{eqA-9}
\end{eqnarray}
One can finally rewrite Eq.~(\ref{eqA-3}) with Eq.~(\ref{eqA-9}) and the Lorentz invariant relation $f_{o}(E_{o}) = f_{c}(E_{c})$ as
\begin{equation}
N_{e}^{o} = \gamma_{o} \gamma_{c} \left(1 - \vec{\beta}_{o} \cdot \vec{\beta}_{c} \right)  N_{e}^{c}  \, .
\label{eqA-10}
\end{equation}
In deriving Eq.~(\ref{eqA-10}), the terms in the $\{ \, \}$ brackets in Eq.~(\ref{eqA-9}) were dropped by the integral over the solid angles, because $f_{c}(E_{c})$ is isotropic.  Finally, it is clear from Eq.~(\ref{eqA-10}) that one obtains $N_{e}^{o} = N_{e}^{c}$ in the limit of $\vec{\beta}_{o}=\vec{\beta}_{c}$ and $N_{e}^{o}=N_{e}=\gamma_{c} N_{e}^{c}$ in the limit of $\vec{\beta}_{o}=0$.

\section[]{Fokker-Planck Expansion Approximation For the Thermal Electron distribution}

\subsection{Rate equation in the CG frame}

In \citet{noza09b}, a formal solution was shown for the rate equation of the leading-order SZ effect by introducing an operator $\mathcal{O}(D)$=$\int_{-\infty}^{+\infty}ds P_{1}(s) e^{s D}-1$.  In the present appndix, we extend the operator representation formalism to the rate equation of Eq.~(\ref{eq3-1-10}).  Let us first introduce an operator $\mathcal{O}_{\ell}(D_{z})$ by
\begin{equation}
\mathcal{O}_{\ell}(D_{z}) \equiv \int_{-\infty}^{+\infty} d s_{c} \, P_{\ell,c}(s_{c}) \, e^{s_{c} D_{z}} \, - \, 1  \, .
\label{eqB-1-1}
\end{equation}
Then, Eq.~(\ref{eq3-1-10}) can be rewritten as follows:
\begin{eqnarray}
\frac{d n_{c}(x_{c})}{d \tau_{c}}  = \left[ \mathcal{O}_{0}(D_{x_{c}}) + \frac{1}{6} \beta_{c}^{2} \, \mathcal{O}_{0}(D_{x_{c}}) \, D_{x_{c}} (D_{x_{c}} + 2) \right.  \nonumber  \\
&& \hspace{-65mm}
+ \, \beta_{c} P_{1}(\mu_{c}^{c}) \, \mathcal{O}_{1}(D_{x_{c}}) \, D_{x_{c}} \nonumber  \\
&& \hspace{-65mm}
+ \, \frac{1}{3} \beta_{c}^{2} P_{2}(\mu_{c}^{c}) \, \mathcal{O}_{2}(D_{x_{c}}) \, D_{x_{c}} (D_{x_{c}} - 1) \bigg] \, n(x_{c}) \, .
\label{eqB-1-2}
\end{eqnarray}

  We now derive the expressions for the operators $\mathcal{O}_{\ell}(D_{z})$ in the Fokker-Planck expansion approximation for the thermal electron distribution.  They are as follows:
\begin{equation}
\mathcal{O}_{0}(D_{z}) = \sum_{n=0}^{5} \mathcal{O}_{0,n} \theta_{e}^{n}   \, ,
\label{eqB-1-3}
\end{equation}
\begin{equation}
\mathcal{O}_{0,0} = 0  \, ,
\label{eqB-1-4}
\end{equation}
\begin{equation}
{\cal O}_{0,1} = \Delta_{z}  \, ,
\label{eqB-1-5}
\end{equation}
\begin{equation}
{\cal O}_{0,2} = -\frac{3}{10} \Delta_{z} + \frac{7}{10} \Delta_{z}^{2} \, ,
\label{eqB-1-6}
\end{equation}
\begin{equation}
{\cal O}_{0,3} = -\frac{31}{120} \Delta_{z} - \frac{14}{15} \Delta_{z}^{2} + \frac{11}{30} \Delta_{z}^{3} \, ,
\label{eqB-1-7}
\end{equation}
\begin{equation}
{\cal O}_{0,4} = \frac{23}{56} \Delta_{z} + \frac{23}{21} \Delta_{z}^{2} - \frac{431}{420} \Delta_{z}^{3} + \frac{16}{105} \Delta_{z}^{4} \, ,
\label{eqB-1-8}
\end{equation}
\begin{equation}
{\cal O}_{0,5} = \frac{1457}{896} \Delta_{z} - \frac{309}{140} \Delta_{z}^{2} + \frac{839}{336} \Delta_{z}^{3} - \frac{74}{105} \Delta_{z}^{4} + \frac{11}{210} \Delta_{z}^{5} \, ,
\label{eqB-1-9}
\end{equation}
\begin{equation}
\mathcal{O}_{1}(D_{z}) = \sum_{n=0}^{5} \mathcal{O}_{1,n} \theta_{e}^{n}   \, ,
\label{eqB-1-10}
\end{equation}
\begin{equation}
\mathcal{O}_{1,0} = -1  \, ,
\label{eqB-1-11}
\end{equation}
\begin{equation}
\mathcal{O}_{1,1} = -\frac{2}{5} -\frac{2}{5} \Delta_{z}  \, ,
\label{eqB-1-12}
\end{equation}
\begin{equation}
\mathcal{O}_{1,2} = -\frac{1}{5} +\frac{2}{5} \Delta_{z} - \frac{2}{5} \Delta_{z}^{2}  \, ,
\label{eqB-1-13}
\end{equation}
\begin{eqnarray}
\mathcal{O}_{1,3} = \frac{407}{140} -\frac{233}{140} \Delta_{z} + \frac{34}{35} \Delta_{z}^{2} - \frac{17}{70} \Delta_{z}^{3}  \, ,
\label{eqB-1-14}
\end{eqnarray}
\begin{eqnarray}
\mathcal{O}_{1,4} = -\frac{363}{28} +\frac{703}{84} \Delta_{z} - \frac{67}{21} \Delta_{z}^{2} + \frac{409}{420} \Delta_{z}^{3} - \frac{23}{210} \Delta_{z}^{4}  \,  ,
\label{eqB-1-15}
\end{eqnarray}
\begin{eqnarray}
\mathcal{O}_{1,5} = \frac{23363}{448} -\frac{160165}{4032} \Delta_{z} + \frac{3547}{252} \Delta_{z}^{2} - \frac{2143}{560} \Delta_{z}^{3}  \nonumber  \\
&& \hspace{-56mm}
+ \frac{29}{45} \Delta_{z}^{4}  - \frac{5}{126} \Delta_{z}^{5}  \, ,
\label{eqB-1-16}
\end{eqnarray}
\begin{equation}
\mathcal{O}_{2}(D_{z}) = \sum_{n=0}^{5} \mathcal{O}_{2,n} \theta_{e}^{n}   \, ,
\label{eqB-1-17}
\end{equation}
\begin{equation}
\mathcal{O}_{2,0} = -\frac{9}{10}  \, ,
\label{eqB-1-18}
\end{equation}
\begin{equation}
\mathcal{O}_{2,1} = -\frac{3}{5} +\frac{1}{10} \Delta_{z}  \, ,
\label{eqB-1-19}
\end{equation}
\begin{equation}
\mathcal{O}_{2,2} = \frac{183}{70} -\frac{3}{4} \Delta_{z} + \frac{1}{7} \Delta_{z}^{2}  \, ,
\label{eqB-1-20}
\end{equation}
\begin{eqnarray}
\mathcal{O}_{2,3} = -\frac{429}{40} +\frac{1535}{336} \Delta_{z} - \frac{14}{15} \Delta_{z}^{2} + \frac{23}{210} \Delta_{z}^{3}  \, ,
\label{eqB-1-21}
\end{eqnarray}
\begin{eqnarray}
\mathcal{O}_{2,4} = \frac{2559}{56} -\frac{8287}{336} \Delta_{z} + \frac{17}{3} \Delta_{z}^{2} - \frac{341}{420} \Delta_{z}^{3} + \frac{2}{35} \Delta_{z}^{4} \,  ,
\label{eqB-1-22}
\end{eqnarray}
\begin{eqnarray}
\mathcal{O}_{2,5} = -\frac{26961}{128} +\frac{704183}{5376} \Delta_{z} - \frac{70843}{2100} \Delta_{z}^{2}  + \frac{43901}{8400} \Delta_{z}^{3} \nonumber  \\
&& \hspace{-59mm}
 - \frac{18}{35} \Delta_{z}^{4}  + \frac{4}{175} \Delta_{z}^{5}  \, ,
\label{eqB-1-23}
\end{eqnarray}
\begin{equation}
\Delta_{z} \equiv D_{z}(D_{z}+3) = \frac{1}{z^{2}} \frac{d}{d z} \left( z^{4} \frac{d}{d z} \right)  \, ,
\label{eqB-1-24}
\end{equation}
where $\theta_{e}=k_{B}T_{\rm CMB}/m$ and $\mathcal{O}(\theta_{e}^{6})$ terms were neglected.

  Equations (\ref{eqB-1-3})--(\ref{eqB-1-24}) show that the Fokker-Planck expansion of the Boltzmann equation is expressed by the power series of the diffusion operator $\Delta_{z}$ of the original Kompaneets equation \citep{komp56}.  Note that the expression for $\mathcal{O}_{0}(D_{z})$ agrees with \citet{noza09b}.  The photon number conservation is guaranteed for the operator $\mathcal{O}_{0}(D_{z})$, because it is written as a function of $\Delta_{z}$ alone.  On the other hand, the $\mathcal{O}_{1}(D_{z})$ and $\mathcal{O}_{2}(D_{z})$ operators contain nonzero constant terms.  Therefore, the photon number conservations are not guaranteed for these operators.  Note also that in Eqs.~(\ref{eqB-1-3}), (\ref{eqB-1-10}) and (\ref{eqB-1-17}) the coefficients are the moments of the frequency shift, which agree with the results of \citet{chlu12b}.  Finally, note that the operators $\mathcal{O}_{\ell}(D_{z})$ satisfy the relations
\begin{equation}
\mathcal{O}_{\ell}(D_{z}) = \mathcal{O}_{\ell}(-D_{z}-3)
\label{eqB-1-25}
\end{equation}
for $\ell$=0, 1 and 2, because $\Delta_{z}$ is invariant under the replacement $D_{z} \rightarrow -D_{z}-3$.  Therefore, any functions of $\Delta_{z}$ satisfy the relation of Eq.~(\ref{eqB-1-25}).

\subsection{Rate equation in the CMB frame}

In order to compare the present work with previous works such as \citet{noza98} (denoted as NIK as under) and \citet{chlu12b}, we reexpress Eq.~(\ref{eq4-1-8}) in the operator representation with Eq.~(\ref{eqB-1-1}).  One has
\begin{eqnarray}
\frac{d n(x)}{d \tau} = \Big[ \mathcal{O}_{0}(D_{x})  \nonumber  \\
&& \hspace{-20mm}
 + \frac{1}{3} \beta_{c}^{2} \, \left\{ \mathcal{O}_{0}(D_{x})  - \mathcal{O}_{1}(D_{x}) \right\} \, D_{x} (D_{x} + 3)  \nonumber  \\
&& \hspace{-20mm}
- \, \beta_{c} P_{1}(\mu_{c}) \, \left\{ \mathcal{O}_{0}(D_{x}) (D_{x}+1) - \mathcal{O}_{1}(D_{x}) D_{x} \right\}   \nonumber  \\
&& \hspace{-20mm}
+ \, \frac{1}{3} \beta_{c}^{2} P_{2}(\mu_{c}) \, \left\{ \mathcal{O}_{2}(D_{x}) \, D_{x} (D_{x} - 1) - 2 \, \mathcal{O}_{1}(D_{x}) \, D_{x}^{2}  \right.  \nonumber  \\
&& \hspace{-1mm}
\left. + \mathcal{O}_{0}(D_{x}) \, D_{x} (D_{x} + 1) \right\}  \Big] \, n(x) \, .
\label{eqB-2-1}
\end{eqnarray}

  On the other hand, one can also express the rate equation of NIK (1998) in the operator representation.  One has
\begin{eqnarray}
\frac{d n(x)}{d \tau} =  \frac{1}{\gamma_{c}} \Big[ \mathcal{O}_{0}(D_{x}) + \frac{1}{3} \beta_{c}^{2} \, \mathcal{O}_{{\rm NIK},3}(D_{x})  \nonumber  \\
&& \hspace{-33mm}
- \, \beta_{c} P_{1}(\mu_{c}) \, \mathcal{O}_{{\rm NIK},1}(D_{x}) \nonumber \\
&& \hspace{-33mm}
+ \, \frac{1}{3} \beta_{c}^{2} P_{2}(\mu_{c}) \, \mathcal{O}_{{\rm NIK},2}(D_{x}) \Big] n(x)  \, .
\label{eqB-2-2}
\end{eqnarray}
In Eq.~(\ref{eqB-2-2}), the factor $1/\gamma_{c}$ was multiplied to the rate equation of NIK (1998) in order to be consistent with the definition of the rate equation of Eq.~(\ref{eqB-2-1}).  The factor agrees with the comment indicated by \citet{chlu12b}.  The operators $\mathcal{O}_{{\rm NIK},\ell}(D_{z})$ are calculated by Eqs.~(14)--(17) of NIK (1998).  They are
\begin{equation}
\mathcal{O}_{{\rm NIK},1}(D_{z}) = \sum_{n=0}^{4} \mathcal{O}_{1,n} \theta_{e}^{n}   \, ,
\label{eqB-2-3}
\end{equation}
\begin{equation}
\mathcal{O}_{1,0} = D_{z}  \, ,
\label{eqB-2-4}
\end{equation}
\begin{equation}
\mathcal{O}_{1,1} = \frac{17}{5} D_{z} + \frac{26}{5} D_{z}^{2} + \frac{7}{5} D_{z}^{3}  \, ,
\label{eqB-2-5}
\end{equation}
\begin{equation}
\mathcal{O}_{1,2} = -\frac{7}{10} D_{z} + \frac{39}{10} D_{z}^{2} + \frac{67}{5} D_{z}^{3} + \frac{73}{10} D_{z}^{4} + \frac{11}{10} D_{z}^{5}  \, ,
\label{eqB-2-6}
\end{equation}
\begin{eqnarray}
\mathcal{O}_{1,3} = -\frac{1031}{280} D_{z} - \frac{373}{84} D_{z}^{2} - \frac{9607}{840} D_{z}^{3} + \frac{2939}{210} D_{z}^{4}
\nonumber  \\
&&\hspace{-66mm}
+ \frac{3749}{210} D_{z}^{5} + \frac{1229}{210} D_{z}^{6} + \frac{64}{105} D_{z}^{7}  \, ,
\label{eqB-2-7}
\end{eqnarray}
\begin{eqnarray}
\mathcal{O}_{1,4} =  \frac{795}{56} D_{z} - \frac{381}{28} D_{z}^{2} + \frac{7961}{840} D_{z}^{3} - \frac{17873}{420} D_{z}^{4}  \nonumber  \\
&&\hspace{-69mm}
- \frac{2979}{140} D_{z}^{5} + \frac{1469}{84} D_{z}^{6} + \frac{489}{35} D_{z}^{7} + \frac{346}{105} D_{z}^{8} + \frac{11}{42} D_{z}^{9}  \, ,
\label{eqB-2-8}
\end{eqnarray}
\begin{equation}
\mathcal{O}_{{\rm NIK},2}(D_{z}) = \sum_{n=0}^{1} \mathcal{O}_{2,n} \theta_{e}^{n}   \, ,
\label{eqB-2-9}
\end{equation}
\begin{equation}
\mathcal{O}_{2,0} = \frac{9}{10} D_{z} + \frac{11}{10} D_{z}^{2}  \, ,
\label{eqB-2-10}
\end{equation}
\begin{equation}
\mathcal{O}_{2,1} = \frac{3}{5} D_{z} + \frac{29}{10} D_{z}^{2} + \frac{33}{5} D_{z}^{3} + \frac{19}{10} D_{z}^{4}  \, ,
\label{eqB-2-11}
\end{equation}
\begin{equation}
\mathcal{O}_{{\rm NIK},3}(D_{z}) = \sum_{n=0}^{1} \mathcal{O}_{3,n} \theta_{e}^{n}   \, ,
\label{eqB-2-12}
\end{equation}
\begin{equation}
\mathcal{O}_{3,0} = D_{z} (D_{z} + 3)  \, ,
\label{eqB-2-13}
\end{equation}
\begin{equation}
\mathcal{O}_{3,1} = D_{z} (D_{z} + 3) \left( \frac{19}{10} + \frac{21}{5} D_{z} + \frac{7}{5} D_{z}^{2} \right)  \,  .
\label{eqB-2-14}
\end{equation}

  Comparing Eqs.~(\ref{eqB-2-3})--(\ref{eqB-2-14}) with Eqs.~(\ref{eqB-1-3})--(\ref{eqB-1-23}), one finds that two calculations agree for the $\beta_{c}$ term and $\beta_{c}^{2} P_{2}(\mu_{c})$ term.  On the other hand, as for the $\mathcal{O}_{\rm NIK,3}(D_{z})$ operator, there is a difference in Eq.~(\ref{eqB-2-14}) by amount of $3/2D_{z}(D_{z}+3)$.  However, this term exactly cancels with the $\beta_{c}^{2}$ term in the $1/\gamma_{c} \, \mathcal{O}_{0}(D_{z})$ term.  Therefore, the two calculations agree completely.  This clarifies the relation of the calculations between the CMB and CG frames.

\section{Properties of the redistribution functions}

We study properties of the redistribution functions.  As already indicated in Section 3, $P_{\ell,c}(s_{c},\beta)$ and $P_{\ell,c}(s_{c})$ satisfy the following relations:
\begin{equation}
P_{\ell,c}(s_{c}, \beta) = e^{3 s_{c}} \, P_{\ell,c}(-s_{c}, \beta)
\label{eqC-1-1}
\end{equation}
and
\begin{equation}
P_{\ell,c}(s_{c}) = e^{3 s_{c}} \, P_{\ell,c}(-s_{c})  \, .
\label{eqC-1-2}
\end{equation}
In Eqs.~(\ref{eq6-1-14})--(\ref{eq6-1-19}), all functions in the [ ] brackets are even functions of $s_{c}$.  Therefore, a multiplicative function $e^{3s_{c}/2}$ in $P_{\ell \ell^{\prime},c}(s_{c},\beta)$ gives the property of Eq.~(\ref{eqC-1-1}).  Equation (\ref{eqC-1-2}) is obtained with Eq.~(\ref{eq6-1-10}).

  Now, we study the behavior of the function $P_{\ell,c}(s_{c},\beta)$ at specific values of $\beta$.  First, one has at $\beta=\beta_{\rm min}$,
\begin{equation}
P_{\ell,c}(s_{c},\beta_{\rm min}) = 0  \, ,
\label{eqC-1-3}
\end{equation}
for $\ell$=0, 1 and 2.  This can be verified with the following identity relations:
\begin{equation}
{\rm cosh} \frac{s_{c}}{2} - \frac{1}{\beta_{\rm min}} \, {\rm sinh} \frac{|s_{c}|}{2} = 0
\label{eqC-1-4}
\end{equation}
and $\lambda_{\beta_{\rm min}} - |s_{c}| = 0$.  Then, one has at $\beta$=1 as follows:
\begin{equation}
\gamma^{2} \, P_{0,c}(s_{c}, 1) = \frac{3}{4} e^{3 s_{c}/2} \, e^{-|s_{c}|/2}  \,  ,
\label{eqC-1-5}
\end{equation}
\begin{equation}
\gamma^{2} \, P_{1,c}(s_{c}, 1) = \frac{3}{4} e^{3 s_{c}/2} \, \frac{1}{3} \, e^{-3|s_{c}|/2}  \,  ,
\label{eqC-1-6}
\end{equation}
\begin{equation}
\gamma^{2} \, P_{2,c}(s_{c}, 1) = \frac{3}{4} e^{3 s_{c}/2} \, \frac{1}{5} \, e^{-5|s_{c}|/2}  \,  .
\label{eqC-1-7}
\end{equation}
These properties are useful for checking the numerical accuracy of the numerical calculations.

\end{document}